\documentclass[12pt]{article}
\usepackage{geometry}                
\usepackage{graphicx}
\usepackage{amssymb}
\usepackage{epstopdf}
\def\gbar{\bar g}
\def\bnabla{\bar\nabla}
\def\dbar{\bar D}
\def\kmax{k_{max}}
\def\lmax{l_{max}}

\begin{document}

\begin{titlepage}
\vfill
\begin{flushright}
\end{flushright}

\vfill
\begin{center}
\baselineskip=16pt
{\Large\bf An Extended Kerr-Schild Ansatz}
\vskip 1.0cm
{\large {\sl }}
\vskip 10.mm
{\bf Benjamin Ett\footnote{bett@physics.umass.edu} and David Kastor\footnote{kastor@physics.umass.edu}} \\
\vskip 1cm
{
Department of Physics\\
University of Massachusetts\\
Amherst, MA 01003
     	     }
\vspace{6pt}
\end{center}
\vskip 0.5in
\par
\begin{center}
{\bf Abstract}
 \end{center}
\begin{quote}
We present an analysis of the vacuum Einstein equations for a recently proposed extension of the Kerr-Schild ansatz that includes a spacelike vector field as well as the usual Kerr-Schild null vector.
We show that many, although not all, of the simplifications that occur in the Kerr-Schild case continue to hold for the extended ansatz.
In particular, we find a simple set of sufficient conditions on the vectors such that the vacuum  field equations truncate beyond quadratic order in an expansion around a general vacuum background solution.
We extend our analysis to the electrovac case with a related ansatz for the gauge field.
\vfill
\vskip 2.mm
\end{quote}
\end{titlepage}


\section{Introduction}

The Kerr-Schild (KS) ansatz \cite{Kerr-Schild}\cite{Debney:1969zz} yields remarkable simplifications of the Einstein equations and has long proved to be a powerful tool in the search for new black hole solutions.
One takes the spacetime metric to have the form
\begin{equation}\label{kerrschild}
g_{ab} = \bar g_{ab} + \lambda h_{ab} ,\qquad
h_{ab} = Hk_ak_b
\end{equation}
where  $\bar g_{ab}$ is a  background metric, $k^a$ is null with respect to the background metric (with $k_a\equiv \bar g_{ab}k^b$), $H$ is a function and the constant $\lambda$ is inserted  for convenience.  To solve the vacuum Einstein equations, the background metric is taken to be Ricci flat.
One can then analyze the Einstein equations order by order in $\lambda$ in an expansion around the background.
A drastic reduction in complexity  comes about because the inverse metric truncates beyond first order in $\lambda$, {\it i.e.}  it is given exactly by 
\begin{equation}\label{ksinverse}
g^{ab} = \bar g^{ab} - \lambda h^{ab}
\end{equation}
with $h^{ab}=\gbar^{ac}\gbar^{bd}h_{cd}$.
Further computation shows that, if the null vector is tangent to a geodesic congruence of the background metric, then the Ricci tensor $R^a{}_b$ of the KS metric
also truncates beyond linear order in $\lambda$ \cite{Gurses}.  These results can also be generalized to non-vacuum cases \cite{Dereli:1986cm}.

The KS ansatz has, in particular, served as the key for finding many higher dimensional black hole solutions.  Myers and Perry \cite{Myers:1986un} made use of it to find neutral, rotating black holes solutions in $D>4$, a task that would very likely have proved intractable otherwise.  The general higher dimensional (A)dS neutral rotating black holes were similarly found by Gibbons et. al.  \cite{Gibbons:2004uw} starting from (A)dS background metrics.

Nevertheless, one could make a long list of potentially interesting black hole 
solutions that have not so far been found via  the KS ansatz (or by any other method).   
Candidates for this list would include the rotating, charged black holes of Einstein-Maxwell theory for $D>4$, vacuum black holes with non-spherical event horizon topology beyond $D=5$ ({\it e.g.} such as those discussed in \cite{Emparan:2009cs}), as well as black branes and rotating black holes in Lovelock gravity theories (beyond the special cases found in \cite{Kastor:2006vw,Giribet:2006ec} and \cite{Anabalon:2009kq} respectively).

Moreover, there are known higher dimensional black hole solutions that cannot be written in KS form, in particular the $5$-dimensional black ring \cite{Emparan:2001wn}.   This may be seen in the following way.  The KS ansatz was originally put forth in the context of algebraically special spacetimes.  In four dimensions, with a flat background metric, the null vector $k^a$ in a vacuum KS spacetime is necessarily a repeated principal null vector of the Weyl tensor \cite{Gurses}.  
In higher dimensions, it was shown in reference \cite{Ortaggio:2008iq} that the Weyl tensor of vacuum KS spacetimes is always of Type $II$, or more, algebraically special, within the classification scheme of reference \cite{Coley:2004jv}.    On the other hand, it was shown in \cite{Pravda:2005kp} that the black ring is only Type $I_i$, and therefore cannot be of KS form.

It seems reasonable to ask whether it might be possible to extend the KS ansatz  in a way that might {\it e.g.} encompass the $D=5$ black ring, or allow one to find further new  black hole solutions  such as those  listed above.  One possible extension was suggested recently in  reference \cite{Aliev:2008bh}.    The authors showed that the charged, rotating black holes of minimal, gauged $D=5$ supergravity, originally found in reference \cite{Chong:2005hr} and known as the CCLP spacetimes, may be rewritten in a form similar to  (\ref{kerrschild}), with $g_{ab} = \bar g_{ab} + \lambda h_{ab}$ and $\bar g_{ab}$  a flat background metric, but now with
\begin{equation}\label{extended}
h_{ab} =  H k_ak_b + K (k_al_b +l_ak_b).
\end{equation}
Here $k^a$ is again a null vector, $H$ and $K$ are functions, the vector $l^a$ is spacelike and  orthogonal to $k^a$ with respect to $\bar g_{ab}$,  and we define $k_{a}\equiv\bar  g_{ab} k^b$ and $l_{a}\equiv\bar  g_{ab} l^b$.   We will call metrics of this general form  extended Kerr-Schild or xKS metrics.

Another indication of the usefulness of the xKS ansatz comes from  considering higher dimensional pp-waves, which are defined by  having a covariantly constant null vector (and hence a null Killing field).  These spacetimes have long been of interest as exact string backgrounds \cite{Horowitz:1989bv}.    It is known that in $D=4$, all pp-wave spacetimes can be cast into Kerr-Schild form (see \cite{Stephani:2003tm}).  However, as discussed in \cite{Ortaggio:2008iq}, examples of pp-wave spacetimes are known in higher dimensions that have Weyl types \cite{Coley:2004jv} that are not compatible with those of Kerr-Schild spacetimes.  Therefore, not all higher dimensional pp-waves can be cast in Kerr-Schild form.  On the other hand, the  particular example of a non-Kerr-Schild pp-wave given in \cite{Ortaggio:2008iq} is of  xKS form, and one may speculate that perhaps all higher dimensional pp-waves can be cast in  xKS form.

The main focus of this paper will be an analysis of the vacuum Einstein equations for xKS metrics.  As an indication of the simplifications we will find, consider the inverse of an xKS metric.  An elementary calculation shows that this truncates beyond second order in $\lambda$, being given exactly by
\begin{equation}\label{xksinverse}
g^{ab} = \bar g^{ab} - \lambda h^{ab} + \lambda^2 h^{ac}h_c{}^b.
\end{equation}
Recall that the truncation of the inverse metric beyond linear order in the Kerr-Schild case led to a similar truncation of the Ricci tensor $R^a{}_b$ 
beyond linear order.  Our main task below is to discover the degree of simplification of the Ricci tensor that occurs in the xKS case.  We will see that for $k^a$  geodesic and $l^a$ also satisfying a certain condition with respect to the background metric, that the  Ricci tensor $R^a{}_b$ will truncate beyond second order in $\lambda$.  The  vacuum Einstein equations then reduce  to a set of differential equations that are quadratic in  $h_{ab}$.

The paper proceeds as follows.  In section (\ref{CCLP}) to further set the context for our work,  we present the xKS form \cite{Aliev:2008bh} of the CCLP spacetimes \cite{Chong:2005hr} (in the limit of vanishing cosmological constant) in Cartesian coordinates that highlight the sense in which they generalize the $D=5$ Myers-Perry metrics \cite{Myers:1986un}.  In section (\ref{basics}) we introduce the basic geometrical tools we will use in computing the Ricci tensor for xKS spacetimes.  In section (\ref{vacuumKS}) we reproduce the analysis in the KS case as a warm-up and as a basis of comparison for the xKS case.  Section (\ref{vacuumxks}) contains our main results on the simplification of the vacuum Einstein equations for xKS metrics.  In section (\ref{stress-energy}) we study the implications of adding an electromagnetic field with a specific form for the gauge potential.  Section (\ref{conclusions}) contains some concluding remarks and suggestions for further research.

After our work was largely complete, reference \cite{bonanos} was brought to our attention, which analyzes a closely related extension of the Kerr-Schild ansatz.  
The contents of  \cite{bonanos} are largely complementary to those of the present paper, being in certain respects both more general and more limited.
The metric ansatz in  \cite{bonanos} is more general in that it allows a variety of possible signs for the norms of the two orthogonal vectors 
$k^a$ and $l^a$.  On the other hand, the material presented in  \cite{bonanos} is more limited in part because it treats only $D=4$ and flat background metrics.  More importantly, however,  our specific case of interest, $k^a$ null and $l^a$ spacelike, is explored in less depth and in particular our main result, a simple, sufficient condition for the truncation of $R^a{}_b$ beyond second order in $\lambda$, is not obtained.

\section{The xKS form of CCLP spacetimes}\label{CCLP}

Our work was motivated by the observation \cite{Aliev:2008bh} that the charged rotating black holes of minimal $D=5$ supergravity \cite{Chong:2005hr}, known as the CCLP solutions, may be written in the extended Kerr-Schild form (\ref{extended}).   In \cite{Aliev:2008bh} the metrics are presented in a type of spheroidal coordinates.  Following a sequence of  steps given in the Appendix, we have transformed them into Cartesian coordinates.  The background  is then simply $5$-dimensional Minkowski spacetime $d\bar s^2 = -d\tau^2 + dx^2 + dy^2 +dw^2 + dz^2$,  while the vector fields 
$k^a$ and $l^a$ are then given by
\begin{eqnarray}\label{cartesian}
k_adx^a &=& d\tau - { r(xdx+ydy) +a (xdy-ydx)\over r^2 +a^2}
- { r(wdw+zdz) +b(wdz-zdw)\over r^2 +b^2} \\
l_adx^a &=&   {b\left( a (xdx+ydy) - r(xdy-ydx)\right)\over r (r^2+a^2)}
+  {a\left( b(wdw+zdx)-r(wdz-zdw)\right)\over r(r^2 +b^2)}.
\end{eqnarray}
The functions $H$ and $K$ in (\ref{extended}) are given by
\begin{equation}\label{metricfunctions}
H = {2m\over\Sigma}-{Q^2\over \Sigma^2},\qquad K = {Q\over\Sigma}.
\end{equation}
Here $\Sigma = r^2+a^2\cos^2\theta +b^2\sin^2\theta$, $r$ is the spheroidal radial coordinate satisfying
\begin{equation}\label{spheroidal}
{x^2 +y^2\over r^2 + a^2} + {w^2 +z^2\over r^2 + b^2} = 1
\end{equation}
and the angle $\theta$ is defined in equation (\ref{cartesiancoords})  in the Appendix.
The $1$-form gauge potential is given by $A= ( \sqrt{3}Q/2\Sigma)k$.

The Cartesian form of the metric facilitates a comparison of the metric with the general odd-dimensional form of the Myers-Perry uncharged rotating black holes \cite{Myers:1986un}.  The vector $k^a$ is identical to that which appears in the $D=5$ Myers-Perry uncharged rotating black holes \cite{Myers:1986un}.
Like the null vector $k^a$, the spacelike vector $l^a$ is independent of the mass  $m$ and charge $Q$ of the spacetime.  The vector $l^a$ can also be seen to be separately orthogonal to $k^a$ in both the $xy$ and $wz$-planes.  This Cartesian form of the metric should be useful in searching for higher dimensional generalizations of the CCLP spacetimes \cite{ek-in-progress}.

We also note some further properties of the vectors $k^a$ and $l^a$ in the CCLP spacetimes.  The null vector $k^a$ satisfies $k^a\bar\nabla_a k^b=0$, where 
$\bnabla_a$ is the covariant derivative operator for the background metric. The vector $k^a$ is thus tangent to a congruence of affinely parameterized null geodesics.  This property is indeed central to the Kerr-Schild construction of the Myers-Perry spacetimes \cite{Myers:1986un}.  In combination, the vectors $k^a$ and $l^a$ can also be shown to satisfy
\begin{equation}
k^b(\partial_b l_a-\partial_a l_b) = 0,\qquad l^b(\partial_b k_a-\partial_a k_b) = 0.
\end{equation}
These same statements would, of course, hold with respect to the background covariant derivative operator $\bnabla_a$ as well.  These equations also imply the 
relation
\begin{equation}
k^a\bnabla_a l^b = -l^a\bnabla_a k^b.
\end{equation}
between the covariant derivative of each vector field along the other.

\section{Computational Basics}\label{basics}

The curvatures of  KS and xKS metrics as well as other useful quantities may be computed in terms of the curvature of the background metric $\bar g_{ab}$  and the background covariant derivatives of the vectors $k^a$ and $l^a$ (in the xKS case).  In this section we present the basic formalism that goes into these calculations.   Let $\bnabla_a$ denote the covariant derivative operator compatible with the background metric $\bar g_{ab}$.  The action covariant of the full covariant derivative of the (x)KS metric on a vector can then be written as
$\nabla_a v^b = \bnabla_a v^b + C_{ac}^b v^c$
with the tensor $C_{ab}^c$ given by
\begin{equation}
C_{ab}^c = {\lambda\over 2}g^{cd}\left(\bnabla_a h_{bd} + \bnabla_b h_{ad} -\bnabla_d h_{ab}\right).
\end{equation}
It is easily checked that the determinant of the (x)KS metric is unchanged from its background value and hence the quantity $C_{ab}^b$ vanishes.  The Ricci tensor of the (x)KS metric is then given by
$R_{ab} = \bnabla_c C_{ab}^c - C_{ac}^d C_{bd}^c.$

We can  write the connection coefficients and  the Ricci tensor as a sum over contributions at different orders in $\lambda$.  Given that the inverse 
KS and xKS metrics truncate beyond orders $\lambda$ and $\lambda^2$ respectively,  the connection coefficients can be written as
\begin{equation}\label{connection}
C_{ab}^c = \sum_{k=1}^{\kmax} \lambda^k C_{ab}^{(k)c}
\end{equation}
where $\kmax =2$ in the KS case and $\kmax=3$ in the xKS case.
The Ricci tensor contains terms quadratic in the connection coefficients and hence has an expansion   
\begin{equation}
R_{ab} = \sum_{l=1}^{\lmax} \lambda^l R_{ab}^{(l)}.
\end{equation}
going out to order $\lmax = 2\kmax$ in $\lambda$.
We will also find it useful to consider the expansion for the Ricci tensor with indices in mixed position $R^a{}_b=g^{ac}R_{cb}$.  Because of the extra factor of the inverse metric, this has an expansion in powers $\lambda^n$ that goes out to order $n_{max}=5$ for the KS case and to order $n_{max}=8$ in  the xKS case.  The coefficients $R^{(n)a}{}_b$ in the expansion of $R^a{}_b$ are simply related to the coefficients in the expansion of $R_{ab}$, with for example
\begin{equation}
R^{(2)a}{}_b = \gbar^{ac}R^{(2)}_{cb} - h^{ac}R^{(1)}_{cb}+h^{ad}h_d{}^c\bar R_{cb}.
\end{equation}
However, they organize the expansion in a different way that turns out to simplify the analysis of the extended Kerr-Schild case below.

Before turning to explicit computations of the Ricci tensor for (x)KS metrics, we consider the important case when the vector $k^a$ is geodesic with respect to the background metric.  
Assuming an affine parameterization we then have $k^a\bnabla_a k^b=0$.  It is well known that for KS metrics, the vector $k^a$ is then also geodesic with respect to the full metric.  This turns out to be true in the xKS case as well.
One has $k^a\nabla_a k^b = k^a\bnabla_a k^b + C^b_{ac}k^ak^c$
and one can check that the quantity $C^b_{ac}k^ak^c$ vanishes for xKS metrics.  Moreover one can show that the expansion, shear and twist of $k^a$ are the same in the xKS metric as in the background.

\section{Vacuum Einstein equations for Kerr-Schild spacetimes}\label{vacuumKS}

The analysis of the vacuum Einstein equations, $R_{ab}=0$, for Kerr-Schild metrics will serve as a model for the analysis in the extended Kerr-Schild case below
(see {\it e.g.} references \cite{xanthopoulos,Chandrasekhar:1985kt} for a similar treatment of the KS case).
We start by rescaling the null vector in (\ref{kerrschild}) to absorb the function $H$ and work with the KS ansatz in the form
$g_{ab} = \bar g_{ab} +\lambda h_{ab}$ with $h_{ab}=k_ak_b$.
We assume that the background metric also satisfies the vacuum Einstein equations, so that $\bar R_{ab}=0$.
The coefficients in $C_{ab}^{(n)c}$ the expansion of the  connection coefficients (\ref{connection}) can be written as
\begin{equation}
C_{ab}^{(1)c} = {1\over 2}\left(\bnabla_a h_b{}^c +\bnabla_b h_a{}^c - \bnabla^ch_{ab}\right),\qquad
C_{ab}^{(2)c} = {1\over 2} k^c \dbar k_ak_b
\end{equation}
where $\bar D = k^a\bnabla_a$ is the background covariant derivative taken along the null vector $k^a$.  

We initially consider the expansion of the Ricci tensor $R_{ab}$ with both of its indices down which goes out to order $\lambda^4$.  Computation shows that the fourth order contribution $R^{(4)}_{ab}$ vanishes identically.  Further progress it is facilitated by considering the contracted equation  $R_{ab}k^ak^b=0$.  One finds that $R^{(3)}_{ab} k^ak^b$ and  $R^{(2)}_{ab} k^ak^b$ vanish identically, while
\begin{equation}
R^{(1)}_{ab}k^ak^b = - (\dbar k_a)\dbar k^a.
\end{equation}  
The vacuum Einstein equation then implies that $\dbar k^a$ is a null vector.  Since it is also orthogonal to $k^a$, it follows that the vector $\dbar k^a$ must be parallel to the null vector $k^a$, {\it i.e.} that  $\dbar k^a = \phi k^a$
for some function $\phi$.  This is equivalent to the statement that $k^a$ is tangent to a null geodesic congruence of the background metric.  Assuming this to be the case, it then follows that the contribution to the Ricci tensor at order $\lambda^3$, which is given by $R^{(3)}_{ab} = -{1\over 2}k_ak_b(\dbar k_d)\dbar k^d$, vanishes as well.  
The contribution at order $\lambda^2$ does not vanish automatically for $k^a$ geodesic.  However, one can show the for geodesic $k^a$, it is related to the order $\lambda^1$ according to 
\begin{equation}\label{R2R1}
R^{(2)}_{ab} = k_ak^c R^{(1)}_{cb}.
\end{equation}
Therefore the vacuum field equations will be satisfied if $R^{(1)}_{ab} =0$.   
This establishes that for Kerr-Schild metrics  with a geodesic null vector $k^a$, solving the vacuum field equations reduces to solving the linearized equations in $h_{ab}$ around the background metric.
This is the result that we will seek an analogue of in the extended Kerr-Schild case.

The corresponding analysis in the extended Kerr-Schild case is more lengthy and intricate.  As noted above, one useful calculational strategy is to work with the expansion of the Ricci tensor $R^{a}{}_b$ with its indices in mixed position.  The potential benefits of this strategy are already evident in the KS case.  After assuming that $k^a$ is geodesic, we found the relation (\ref{R2R1}) between the first and second order terms in the expansion of $R_{ab}$.  The equivalent statement in terms of the expansion of $R^a{}_b$ is simply $R^{(2)a}_b=0$.
Overall,  one finds that the quantities $R_{a}^{(n)b}$ vanish identically for $n=3,4,5$ for the Kerr-Schild ansatz, while $R_{a}^{(2)b}$ vanishes after making use of the geodesic condition.  One is then left in a slightly more straightforward manner with the single equation $R_{a}^{(1)b}=0$.  

\section{Vacuum Einstein equations for extended Kerr-Schild spacetimes}\label{vacuumxks}

We will now present a similar analysis of the vacuum Einstein equations for extended Kerr-Schild metrics.  We are interested in seeing  what simplifications will occur and, in particular,  whether the expansion of the Ricci tensor will truncate beyond some relatively low order in $\lambda$.  We are also interested to see whether the vector $l^a$ should be taken to satisfy some condition that complements the geodesic condition for $k^a$.

We begin by rescaling the vectors $k^a$ and $l^a$ in the xKS ansatz to absorb the functions $H$ and $K$ in (\ref{extended}).  Using the same symbols for the rescaled vectors, the xKS ansatz then takes the form
\begin{equation}
g_{ab}=\bar g_{ab} +\lambda h_{ab},\qquad h_{ab} = k_ak_b + k_al_b + l_a k_b
\end{equation}
with the  vectors still assumed to satisfy $k_ak^a = k_al^a =0$.
For the xKS ansatz one has $C_{ab}^c = \sum_{k=1}^{3} \lambda^k C_{ab}^{(k)c}$ and 
$R_{ab} = \sum_{l=1}^{6} \lambda^l R_{ab}^{(l)}$.  Many computations simplify by using the relations between terms of successive order in the expansion for the connection coefficients
\begin{equation}
C_{ab}^{(2)c} =   -h^c{}_d C_{ab}^{(1)d},\qquad C_{ab}^{(3)c}  =   h^c{}_d h^d{}_eC_{ab}^{(1)e} ,
\end{equation}
where the first order term is simply $C_{ab}^{(1)c}  =   {1\over 2}\left(\bnabla_a h_b{}^c +\bnabla_b h_a{}^c - \bnabla^ch_{ab}\right)$.

Proceeding initially with the expansion for the Ricci tensor with both indices down, it follows that $R_{ab}^{(6)}$ vanishes identically.  Considering next the contracted equation $R_{ab}k^ak^b=0$, one finds that $R_{ab}^{(n)}k^ak^b$ with $n=5,4,3$ vanish identically, while at order $\lambda^2$ one finds
\begin{equation}
R_{ab}^{(2)}k^ak^b = -{1\over 4}\alpha_{ab}\alpha^{ab}
\end{equation}
where $\alpha_{ab}=l_a\dbar k_b -l_b \dbar k_a$.  The vacuum equation implies that the anti-symmetric tensor $\alpha_{ab}$ must be null.  Together with the identity $k^a\dbar k_a=0$, this implies that $\dbar k^a$ must have the form
\begin{equation}\label{dbarform}
\dbar k^a = \phi k^a +\eta l^a
\end{equation}
for some functions $\phi$ and $\eta$.  At order $\lambda^1$ one finds
\begin{equation}
R_{ab}^{(1)}k^ak^b =  -\dbar(l_a\dbar k^a) -(\bnabla_c k^c) l_a\dbar k^a
- (\dbar k_a)(\dbar k^a +\dbar l^a +l^b\bnabla_b k^a)
\end{equation}
Substituting the form (\ref{dbarform}) into this result gives
\begin{equation}\label{badequation}
R_{ab}^{(1)}k^ak^b = - \bnabla_c(\eta k^c l_b l^b) - \eta^2 l_b l^b -\eta l^bl^c\bnabla_b k_c .
\end{equation}
It is clear that taking $k^a$ to be tangent to a geodesic of the background ({\it i.e.} taking $\eta=0$) solves $R_{ab}^{(1)}k^ak^b=0$.  However, it is unclear whether 
null vectors $k^a$ satisfying (\ref{dbarform}) with $\eta\ne 0$ are possible.  We will proceed by assuming that $k^a$ is geodesic.  As noted in section (\ref{CCLP}) the null vector field in the CCLP spacetimes satisfies $\dbar k^a=0$.
%

Given the geodesic condition, one can then show that $R_{ab}^{(5)}=0$.  
Continuing on with the calculation of $R_{ab}^{(4)}$, however, proves to be quite cumbersome.  In order to proceed we will alter our approach in two ways.
The first change in strategy is to work instead with the expansion of $R^{a}{}_b$ as discussed above.  The second is to adopt  a simpler, but still equivalent,  form for 
$h_{ab}$.  Given that the null vector $k^a$ is assumed to satisfy the geodesic condition we can rescale it by a function such that the rescaled vector satisfies $\dbar k^a =0$ ({\it i.e.} so that the geodesic congruence to which it is the tangent vector is now affinely parameterized).  Similarly we can rescale the spacelike vector $l^a$ by a function such that the rescaled vector has unit norm with respect to the background metric.  The quantity $h_{ab}$ will now have the form given in (\ref{extended}) for some functions $H$ and $K$, where $k^a$ and $l^a$ now represent the rescaled vectors.  Finally,  we can define a new vector $m^a$ through $m^a = l^a + (H/2K)k^a$.  Because the vectors $k^a$ and $l^a$ are orthogonal, the vector $m^a$ will also have unit norm.
In terms of $m^a$, the tensor $h_{ab}$ then reads
\begin{equation}\label{newform}
h_{ab} = K (k_a m_b + m_a k_b),
\end{equation}
where now $k_ak^a=0$, $m_am^a=1$, $k_am^a=0$ and $\dbar k^a=0$.  This new form for $h_{ab}$ simplifies the calculations considerably.  However, note that it is now harder to take a Kerr-Schild limit of the extended Kerr-Schild calculations.

Given this new form of $h_{ab}$, the quantity
$R^{(4)a}{}_b$ can now be shown to vanish\footnote{We should also check the higher order terms in the expansion of $R^a{}_b$ which goes out to order
 $\lambda^8$. Making use of the results stated above can show that $R^{(n)a}{}_b=0$ for $n=5,\dots,8$.}, while for $R^{(3)a}{}_b$ we obtain the following expression
 \begin{equation}\label{r3}
R^{(3)a}{}_b = {1\over 2}\bnabla_d\left (K^3k_b[k^av^d
- v^a k^d] \right)
-{1\over 2} K^3k^a v^d \bnabla_bk_d
\end{equation}
where\footnote{In the expression for $v_a$, the vector $l^a$ may be replaced by the vector $m^a$ without changing the result for $R^{(3)a}{}_b$.}
\begin{equation}
 v_a = k^b\left\{ (\bnabla_bl_a-\bnabla_al_b) - l^c(\bnabla_b l_c-\bnabla_c l_b  )l_a  \right\} 
 \end{equation}
 %
 A sufficient condition for $R^{(3)a}{}_b$ to vanish is that the vector $v^a$ should satisfy
 \begin{equation}\label{newcondition}
 v^a=\alpha k^a
 \end{equation}
 for some function $\alpha$.  
This condition on $l^a$ may be viewed as a counterpart to the geodesic condition for $k^a$.  It is independent of the metric functions $H$ and $K$, depending only on properties of the vectors $l^a$ and $k^a$ with respect to the background metric.

Given that the condition (\ref{newcondition}) on $l^a$ has only been shown to be sufficient, rather than necessary, for the vanishing of $R^{(3)a}{}_b$, it is interesting to ask whether (\ref{newcondition}) is satisfied by the CCLP spacetimes of section (\ref{CCLP})?  The spacelike vector in the CCLP spacetimes does not have unit norm.  Therefore, let us rescale once again letting $\hat l^a = f l^a$ denote the CCLP spacelike vector having norm $f^2$.   The vector $v^a$ is given in terms of $\hat l^a$ and $f$ by 
\begin{equation}
 v_a = k^b\left\{{1\over f} (\bnabla_b\hat l_a-\bnabla_a\hat l_b) - {1\over f^3}\hat l^c(\bnabla_b \hat l_c-\bnabla_c \hat l_b)  l_a  \right\}
 \end{equation}
the derivatives of $f$ having cancelled out.  As noted in section (\ref{CCLP}), the quantity $k^a(\bnabla_a\hat l_b-\bnabla_b\hat l_a)$ vanishes for the CCLP spacetimes and hence condition (\ref{newcondition}) is satisfied in this case with $\alpha=0$.



We have now established a set of sufficient conditions, the geodesic condition on $k^a$ and the condition (\ref{newcondition}) relating $k^a$ and $l^a$, such that the Ricci tensor with indices in mixed position vanishes beyond quadratic order in 
$\lambda$ for xKS spacetimes.
One is now left to consider only the quantities $R^{(2)a}{}_b$ and $R^{(1)a}{}_b$.  In the KS case, one finds that $R^{(2)a}{}_b$ vanishes as a consequence of the geodesic condition for the null vector $k^a$, and consequently that the vacuum Einstein equations reduce to the equation $R^{(1)a}{}_b=0$, which is linear in $h_{ab}$.  However, this does not appear to happen for the xKS ansatz.  Although we have not shown it definitively, we believe that no manipulations of the expression for  
$R^{(2)a}{}_b$ in the xKS case, using the geodesic condition for $k^a$ in combination with  (\ref{newcondition}), will make $R^{(2)a}{}_b$ vanish.  
A more conclusive argument to this same end will be given in the next section when we consider electrovac xKS spacetimes.

The remaining vacuum equations of motion are then $R^{(1)a}{}_b=0$ and $R^{(2)a}{}_b=0$, with the linear and quadratic contributions to the Ricci tensor given respectively by
\begin{equation}
R^{(1)a}{}_b = {1\over 2}\bnabla_c\left(\bnabla^a h_b{}^c + \bnabla_b h^{ac} -\bnabla^c h^a{}_b\right ) 
\end{equation}
\begin{eqnarray}
R^{(2)a}{}_b  =& -{1\over 2}\bnabla_d\left\{
\bnabla_b(h^{ac}h_c{}^d) +h^d{}_e(\bnabla^ah_b{}^e-\bnabla^e h^a{}_b)
+h^a{}_c(\bnabla^c h_b{}^d - \bnabla^d h_b{}^c)\right \}  \\    \nonumber &
-{1\over 4}(\bnabla^e h^{ac} +\bnabla^ch^{ae}-\bnabla^ah^{ce})
(\bnabla_eh_{bc} -\bnabla_bh_{ce}-\bnabla_ch_{be})
\end{eqnarray}
In summary, we have shown that for $k^a$ and $l^a$ satisfying  $\dbar k^a=0$ and equation (\ref{newcondition}) that  the terms $R^{(n)a}{}_b$ in the expansion of the Ricci tensor vanish for $n=3,\dots,8$.  The vacuum Einstein equations then reduce to the two equations noted above.   Condition (\ref{newcondition}) depends only on properties of the vectors $k^a$ and $l^a$ with respect to the background metric and can be regarded as a counterpart to the geodesic condition on $k^a$.
The vacuum Einstein equations continue to simplify  considerably in the xKS case, although not to the full extent that they do in the original KS case.

\section{Adding stress-energy}\label{stress-energy}

The CCLP spacetimes \cite{Chong:2005hr}  shown to be of xKS form in \cite{Aliev:2008bh} and presented above in section (\ref{CCLP}) are non-vacuum spacetimes.  Therefore, we should consider how the xKS ansatz works in the presence of matter fields.
As in the KS case, a key step in our analysis has been considering the equation $R_{ab}k^ak^b = 0$, which led to the geodesic condition on the null vector $k^a$.  Although one could consider more general cases where the null vector $k^a$ is not geodesic, we will continue to focus on the geodesic case, which implies that the stress-energy tensor should satisfy $T_{ab}k^ak^b = 0$.  We will further restrict our attention to the electromagnetic case, with the stress-energy tensor given by
\begin{equation}\label{stressenergy}
T_a{}^b = F_{ac}F^{bc}-{1\over 4}g_a{}^b F^2
\end{equation}
and assume that the gauge potential is related to the xKS null vector $k^a$ according to
\begin{equation}\label{gauge}
A_a = \sqrt{\lambda}\beta k_a
\end{equation}
where $\beta$ is a function.  This form of the gauge field holds in $D=4$ Kerr-Newman spacetimes, in the KS form of the Reissner-Nordstrom spacetime in any dimension, and also in the CCLP spacetimes \cite{Chong:2005hr} in xKS form \cite{Aliev:2008bh}.  It is easily checked that the condition  $T_{ab}k^ak^b =0$ is satisfied by this ansatz for the gauge potential.  

Let us consider the Kerr-Schild case first.  Given that it is necessary to raise two indices on the field strength tensor using the KS inverse metric (\ref{ksinverse}) in order to compute the components of $T_a{}^b$, there could in principle be contributions out to order $\lambda^3$.  However, calculation shows that this is not the case.  With the ansatz (\ref{gauge}) for the gauge potential, the only non-vanishing contributions to $T^a{}_b$ are linear in $\lambda$.  This is consistent with the reduction in order of the Ricci tensor in mixed form $R^{a}{}_b$.  Had there been a contribution to $T^a{}_b$ at {\it e.g.} order $\lambda^2$, this would have been inconsistent with the vanishing of $R^{(2)a}{}_b$.

Now consider the xKS case.  Given the form of the xKS inverse metric (\ref{xksinverse}), there could in principle be contributions to $T_a{}^b$ out to order
$\lambda^5$.  Computation shows that while the order $\lambda^n$ terms in $T_a{}^b$ vanish for $n=3,4,5$, they will generally be non-zero for both $n=1$ and $n=2$.  This is consistent with our findings above in section (\ref{vacuumxks}), where we found that, in contrast to the KS case, the term $R^{(2)a}{}_b$ does not generally vanish for xKS spacetimes\footnote{It is potentially interesting to note that this same truncation of the stress energy tensor holds if a term $ \sqrt{\lambda}\gamma\, l_a$ is added to the gauge potential (\ref{gauge}) for $\gamma$ an arbitrary function, if $l_a$ is assumed to satisfy condition (\ref{newcondition}) that implies the vanishing of $R^{(3)a}{}_b$.}.

For completeness, we should also consider the gauge field equations of motion.  
For the standard Maxwell Lagrangian, the equations of motion are simply $\nabla_a F^{ab}=0$ and it is straightforward to substitute in the xKS ansatz.
One finds that $F^{ab}=\lambda^{1/2}F^{(1/2)ab} + \lambda^{3/2} F^{(3/2)ab}$ with higher order terms vanishing.
It is natural, however, to also include the contribution to the equations of motion coming from the Chern-Simons term in the action of minimal $D=5$ supergravity that is relevant for the CCLP spacetimes\footnote{Note that the Chern-Simons term does not contribute to the stress energy tensor.}.   
The gauge field equation of motion is then given by
\begin{equation}\label{gaugeeom}
\nabla_a F^{ab} - {1\over 2\sqrt{3}}\epsilon^{bcedf}F_{cd}F_{ef} = 0.
\end{equation}
At this point, however, a conflict arises in the order by order expansion in powers of $\lambda$.
Because $\sqrt{-g}=\sqrt{-\bar g}$ for xKS spacetimes, one can replace the derivative operator in (\ref{gaugeeom}) with the background derivative operator.  The first term in (\ref{gaugeeom}) thus has contributions at orders $\lambda^{1/2}$ and $\lambda^{3/2}$, while the second term is manifestly of order $\lambda^1$.  

We expect that a more subtle analysis would be required in order to properly incorporate the gauge field of minimal $D=5$ supergravity into our analysis.  In hindsight, this is evident from the form of the CCLP spacetimes given in section (\ref{CCLP}).  The gauge field is proportional to the charge, and we may therefore think of $\lambda^{1/2}$ as being proportional to the charge $Q$.  In Reissner-Nordstrom spacetimes or in the four dimensional Kerr-Newman spacetimes, the metric depends only on the square of the charge.  However, the metric function $K$ in (\ref{metricfunctions}) is linear in $Q$.  The CCLP metric then appears to include terms proportional to $\lambda^{1/2}$ as well as $\lambda^1$.   The first term in (\ref{gaugeeom}) is linear in $Q$, while the second term is quadratic.  It can only be solved by virtue of  terms in the metric that are linear in $Q$.

We will not attempt to carry out such a more subtle analysis here.  We note that this issue does not affect our main result in section (\ref{vacuumxks}), the truncation of the Ricci tensor $R^a{}_b$ beyond quadratic order in $\lambda$ for xKS metrics with $k^a$ geodesic and $k^a$ and $l^a$ jointly satisfying the condition (\ref{newcondition}).

\section{Conclusions}\label{conclusions}

In section (\ref{vacuumxks}) we found that the Ricci tensor for xKS ansatz metrics simplified to the extent that the vacuum Einstein equations reduce to equations that are quadratic in $h_{ab}$.  Although this falls short of the simplification that happens in the KS case, we suggest that  the substantial reduction in order that does occur, taken together with the existence of PP-wave and black hole solutions of xKS form, provides strong evidence that xKS metrics are worthy of further attention.

In the KS case, most interesting new solutions have been found not by a broadly based attack on the equations, but rather by generalizing known solutions.  That will even more likely be the case for xKS metrics, where the equations to solve are more complex.  As a start in this direction, we are presently searching \cite{ek-in-progress} for xKS solutions of higher dimensional Einstein-Maxwell-Chern-Simons theories based on the Cartesian form of the  CCLP spacetimes  \cite{Chong:2005hr,Aliev:2008bh} presented in section (\ref{CCLP}).  Additional directions would be asking whether all vacuum PP-waves in $D>4$ can be put in xKS form and studying the Weyl type \cite{Coley:2004jv} of xKS metrics along the lines of the analysis for KS metrics performed in \cite{Ortaggio:2008iq}.

\subsection*{Acknowledgements}

This work is supported by NSF grant PHY-0555304.  The authors thank Marcello Ortaggio for helpful conversations and for pointing out reference \cite{bonanos} and the referees for making many suggestions for improvements to the paper.
DK would like to thank the organizers of the Benasque workshop on ``Gravity - New Perspectives from Strings and Higher Dimensions'' where part of this work was completed.

\appendix

\section{Transforming CCLP spacetimes to Cartesian coordinates}

In this appendix we show how to transform the xKS form of the the $\Lambda=0$ limit of the CCLP metrics given in  \cite{Aliev:2008bh} into the Cartesian coordinates in section (\ref{CCLP}).
The xKS form (\ref{extended}) of the $\Lambda=0$ CCLP spacetimes presented in \cite{Aliev:2008bh} is
\begin{eqnarray}
d\bar s^2 &=& -dt^2 -2 dr(dt-a\sin^2\theta d\phi - b\cos^2\theta d\psi) + \Sigma d\theta^2\nonumber  \\
&&+(r^2 +a^2)\sin^2\theta d\phi^2 +(r^2 +b^2)\cos^2\theta d\psi^2\nonumber\\
k_a\, dx^a &=& dt -a\sin^2\theta  d\phi-b\cos^2\theta d\psi, \\
l_a\, dx^a &=& -b\sin^2\theta d\phi - a\cos^2\theta d\psi\nonumber  
\end{eqnarray}
with the functions $H$ and $K$ and the $1$-form gauge potential $A_adx^a$ as given in section (\ref{CCLP}).
The flat background metric
$\bar g_{ab}$  can be transformed into more standard spheroidal coordinates via a transformation such that 
\begin{equation}
dt=d\tau - dr,\qquad
d\phi = d\varphi - {a\over r^2 +a^2} dr,\qquad
d\psi = d\chi - {b\over r^2 +b^2}dr.
\end{equation}
giving
\begin{eqnarray}
d\bar s ^2& = & -d\tau^2 + {r^2\Sigma\over (r^2 +a^2)(r^2 +b^2)} dr^2 +\Sigma d\theta^2 + (r^2+a^2)\sin^2\theta d\varphi^2 
+ (r^2+b^2)\cos^2\theta d\chi^2 \nonumber\\
k_a\, dx^a &=& d\tau - {r^2\Sigma\over (r^2 +a^2)(r^2 +b^2)} dr - a\sin^2\theta d\varphi -b\cos^2\theta d\chi\\
l_a\, dx^a &=&{ab\Sigma\over (r^2 +a^2)(r^2 +b^2)} dr -b\sin^2\theta d\varphi - a\cos^2\theta d\chi \nonumber
\end{eqnarray}
A further transformation may now be made to Cartesian spatial coordinates via
\begin{eqnarray}\label{cartesiancoords}
x &=& \sqrt{r^2+a^2}\sin\theta\cos\varphi,\qquad
y =  \sqrt{r^2+a^2}\sin\theta\sin\varphi \\
w &=&  \sqrt{r^2+b^2}\cos\theta\cos\chi,\quad
z =  \sqrt{r^2+b^2}\cos\theta\sin\chi \nonumber .
\end{eqnarray}
The spheroidal radial coordinate $r$ satisfies the relation (\ref{spheroidal}).
so that surfaces of large $r$ are approximately spherically symmetric, while as $r$ approaches to zero they degenerate into the product of a disk of radius $a$ in the $xy$-plane with a disk of radius $b$ in the $wz$-plane.  The background metric and the vectors $k^a$ and $l^a$ are then those given in (\ref{cartesian}).

\end{document}